\documentclass[journal]{IEEEtran}
\usepackage[utf8]{inputenc}
\usepackage[T1]{fontenc}

\usepackage{graphicx}
\usepackage{subfig}
\usepackage{float}

\usepackage{xspace}
\usepackage[super]{nth}
\usepackage[noadjust]{cite}
\usepackage[shortlabels, inline]{enumitem}

\newcommand{\eg}{\textit{e.g.,}\xspace}
\newcommand{\ie}{\textit{i.e.,}\xspace}

\usepackage{color, xcolor, colortbl}

\begin{document}
\bstctlcite{IEEEexample:BSTcontrol}
\title{Communication and Computation O-RAN Resource Slicing for URLLC Services Using Deep Reinforcement Learning}

\author{
    \IEEEauthorblockN{
        Abderrahime Filali, \IEEEmembership{Student Member, IEEE},
        Boubakr Nour, \IEEEmembership{Member, IEEE}, \\
		Soumaya Cherkaoui, \IEEEmembership{Senior Member, IEEE}, and
		Abdellatif~Kobbane,~\IEEEmembership{Senior Member,~IEEE}
    }
    
    \thanks{A. Filali, B. Nour and S. Cherkaoui are with the INTERLAB Research Laboratory, Faculty of Engineering, Department of Electrical and Computer Science Engineering, Université of Sherbrooke, Sherbrooke (QC) J1K 2R1, Canada (e-mails: abderrahime.filali@usherbrooke.ca, boubakr.nour@usherbrooke.ca, soumaya.cherkaoui@usherbrooke.ca).}
    
    \thanks{A. Kobbane is with the UM5R ENSIAS, BP 713 Rabat, Morocco (e-mail: abdellatif.kobbane@um5.ac.ma).}
}

\markboth{}{} 

\maketitle

\begin{abstract}
    The evolution of the future beyond-5G/6G networks towards a service-aware network is based on network slicing technology. With network slicing, communication service providers seek to meet all the requirements imposed by the verticals, including ultra-reliable low-latency communication (URLLC) services. In addition, the open radio access network (O-RAN) architecture paves the way for flexible sharing of network resources by introducing more programmability into the RAN.  RAN slicing is an essential part of end-to-end network slicing since it ensures efficient sharing of communication and computation resources. However, due to the stringent requirements of URLLC services and the dynamics of the RAN environment, RAN slicing is challenging. In this article, we propose a two-level RAN slicing approach based on the O-RAN architecture to allocate the communication and computation RAN resources among URLLC end-devices. For each RAN slicing level, we model the resource slicing problem as a single-agent Markov decision process and design a deep reinforcement learning algorithm to solve it. Simulation results demonstrate the efficiency of the proposed approach in meeting the desired quality of service requirements.
\end{abstract}


\IEEEpeerreviewmaketitle

\section{Introduction}
\label{sec:introduction}
Although fifth-generation (5G) standards are not yet fully finalized, the roadmap for sixth-generation (6G) networks is already taking shape due to several industrial and academic research efforts \cite{letaief2021edge}. 6G networks are expected to support more diversified services compared to 5G networks, which should create exciting business opportunities in many vertical sectors \cite{ji2021several} Achieving this requires (i) improving the technologies behind the evolution of 5G, such as network slicing \cite{maule20215g}, and (ii) leveraging machine learning (ML)/artificial intelligence (AI) techniques, such as deep reinforcement learning (DRL) for an efficient management of network resources. In addition, to meeting the requirements of various industries, 6G should not only rely on new enabling technologies but also provide an innovative network architecture beyond current network designs. Open radio access network (O-RAN) is a key component of this architectural transition to more open and intelligent networks \cite{ORAN}. The O-RAN approach sustains the disaggregation between hardware and software to create a multi-supplier RAN solution through open and interoperable protocols and interfaces. The O-RAN specification, which is still compliant with 3GPP standards, introduces the hierarchical RAN Intelligent Controller (RIC), including non-real-time RIC (non-RT) and near-real-time RIC (near-RT) where ML/AI algorithms are integrated to enable RAN programmability.

RAN slicing is a critical component of end-to-end network slicing, as it determines the degree of flexibility network operators have to meet the needs of new verticals. In particular,  ultra-reliable low-latency communications (URLLC) is the foundation for emerging mission-critical applications in 6G networks, such as autonomous driving, industrial IoT, e-health (\eg remote surgery) and mobile or m-health (\eg patient monitoring and virtual reality-assisted care in ambulances). Due to the stringent requirements of these applications, they are expected to rely on multi-access edge computing (MEC) to deliver added value
services to the end users. Therefore, effective management of RAN slicing will rely on the ability to optimally manage communication and computing resources placed at the MEC \cite{filali2020multi}.

Considerable efforts have been devoted to improving the performance of RAN slicing to efficiently offload tasks at the MEC \cite{yang2021ran}, where the RAN resource slicing problem is usually formulated using optimization techniques \cite{su2019resource}. However, due the dynamics of the RAN environment, solving the problem of RAN slicing is 
challenging to solve in polynomial time. 
To  overcome  these  issues, 6G RAN slicing operations will need to be performed  with more  intelligent  resource  allocation  capabilities that achieve delay-efficient performances. Under the O-RAN architecture, RICs can dynamically create multiple RAN slices tailored to URLLC services using ML/AI capabilities such as DRL algorithms. 
 Indeed, non-RT and near-RT RICs can leverage DRL’s excellent learning ability and effectiveness in solving complex and dynamic environment problems, such as the RAN environment, to make optimal RAN slicing decisions for URLLC services \cite{ORAN_2}.

In this work, we are motivated to apply the DRL algorithms within the O-RAN architecture to jointly slice the communication and computation resources at the RAN level for URLLC task offloading operations. Indeed, we propose a two-level RAN slicing approach based on DRL. The first level -- communication slicing level, concerns the allocation of radio resources to end-devices. The second level -- computation slicing level, deals with the allocation of computation resources to end-devices. The contribution of our work is as follows. We first introduce the RAN slicing paradigm and its associated challenges. 
Then, we model each RAN slicing resource level as a single agent Markov decision process. Next, we propose, for each RAN slicing level, a DRL algorithm to solve it. Finally, we illustrate through extensive simulations that the proposed approach exhibits fast convergence and achieves delay-efficient performance.

\section{Unveiling the Curtain: RAN Slicing}
\label{sec:background}
Network slicing is the transformation of a physical network into a set of logical networks on top of a shared infrastructure.
RAN slicing  is a critical part of end-to-end network slicing to enable differentiated traffic processing and isolation. This can be achieved through application-based prioritization of data, resource allocation, and scheduling.

\subsection{ RAN Slicing Efforts}
To date, various efforts have been presented to improve the capacity to deliver URLLC services through RAN slicing. The Third Generation Partnership Project (3GPP) has made significant standardization efforts to define RAN slicing specifications and promote its implementation.
For instance, 3GPP introduced the RAN slicing management framework to manage the life cycle of RAN slices \cite{3gpp_1}. In addition, it provided efficient solutions that allow end-devices to rapidly access a cell and select the desired RAN slices \cite{3gpp_2}. 
A joint RAN slicing framework for communication and computation resources has been developed in \cite{zarandi2021delay}. Communication and computation resources are allocated to RAN slices in order to minimize the delay needed to offload and process time-sensitive users’ tasks. To support a maximum number of RAN slices while meeting their performance requirements, \cite{yang2021ran} proposes to share the radio resource between RAN slices by allocating a fraction of bandwidth that maximizes the access probability to the base station and the energy efficiency of end-devices.

Reinforcement learning (RL)-based RAN slicing approaches have also emerged as practical solutions with low computational complexity and simplified implementation. For instance, \cite{wu2020dynamic} introduces a RL-based framework to dynamically allocate radio spectrum and computation resources to RAN slices. The allocation process considers the delay as the primary QoS metric.
\cite{liu2020edgeslice} employs DRL to design a decentralized RAN resource orchestration system. The latter includes an agent to slice each RAN resource and a central coordinator that manages the resource orchestration between the agents. Each orchestration agent uses DRL to allocate its resources to RAN slices, while the central coordinator ensures SLA requirements.

\subsection{ Open Challenges in RAN Slicing}
Despite the aforementioned solutions, various issues remain open. These issues could be summarized as follows:
    
\begin{itemize}
    \item \textit{Resource Sharing:}
    Efficient resource sharing is a primary objective of RAN slicing. However, when a slice is instantiated, dedicated resources may become unavailable to others. Resources reallocation among slices may further enhance optimizing resources utilization as well as improving the network performance. However, dynamic changes in network load, end-devices mobility, and task distribution make resource reallocation challenging.

    \item \textit{Dynamic Slice Creation/Management:}
    In light of the previous point, optimizing resource allocation is indispensable to maximize verticals’ benefits, where dynamic slice creation/management are critical during the slice lifecycle. To accommodate a maximum amount of service requests with minimum resources, the network operator needs to deploy various dynamic mechanisms to quickly create/manage slices. 
    
    \item \textit{Mobility Management:}
    Today’s users may shift from a network to another while requesting services. Seamless handover and interference management add more challenges to RAN slicing. For instance, it is critical to ensure fast mobility handover for real-time services. The system performance relies on the performance of the handover mechanism. Therefore, there is a need for a slice-oriented mobility management protocol to tackle the mobility issues in RAN slicing.
    
    \item \textit{Algorithmic Aspects:}
    Resource allocation is a challenging problem that often encompasses many parameters and constraints. Different algorithms have been adopted to solve the problem according to its complexity. Exact algorithms can be applied to find optimal solutions for less complex problems, while meta-heuristic algorithms are more efficient when dealing with more complex problems. Therefore, practical 
    resource allocation algorithms are necessary with the ability to reconfigure slice resources based on the dynamic network changes.
\end{itemize}

With O-RAN, the door is now unlocked to enhance RAN slicing and address many of its challenges using the non-RT and near-RT RICs \cite{niknam2020intelligent}. The former handles the heaviest RAN functions, at a time scale > 1s, including robust RAN analytics, control policy design, providing trained AI/ML models and guidance to support near-RT RIC operations. The latter executes critical RAN functions, at a time scale that could be as low as 10ms, to interpret and enforce the received policies from non-RT RIC such as using AI/ML inference to control RAN behavior. Therefore, the interaction between non-RT and near-RT RICs can be used to design and fine-tune efficient AI/ML control algorithms for RAN slicing. 
%
In this work, we leverage the hierarchical RIC features and DRL capabilities to achieve dynamic slice creation and management, as well as efficient resource reallocation and sharing to meet the requirements of URLLC services. Indeed, we propose a DRL-based RAN slicing approach driven by the non-RT RIC and the near-RT RIC. In each resource slicing level, a DRL algorithm can be retrained to dynamically reallocate resources among slices based on changes in network load and task distribution.
Moreover, RICs are used to automate the deployment and monitoring URLLC slices.

\section{Joint Slicing of Communication and Computation RAN Resources}
\label{sec:design}
We describe, in the following, the proposed two-level RAN slicing approach, where the communication and computation RAN resource are jointly sliced and allocated to the end-devices according to their URLLC requirements.

\subsection{System Model}

\begin{figure}[!t]
	\centering
	\includegraphics[width=1\linewidth]{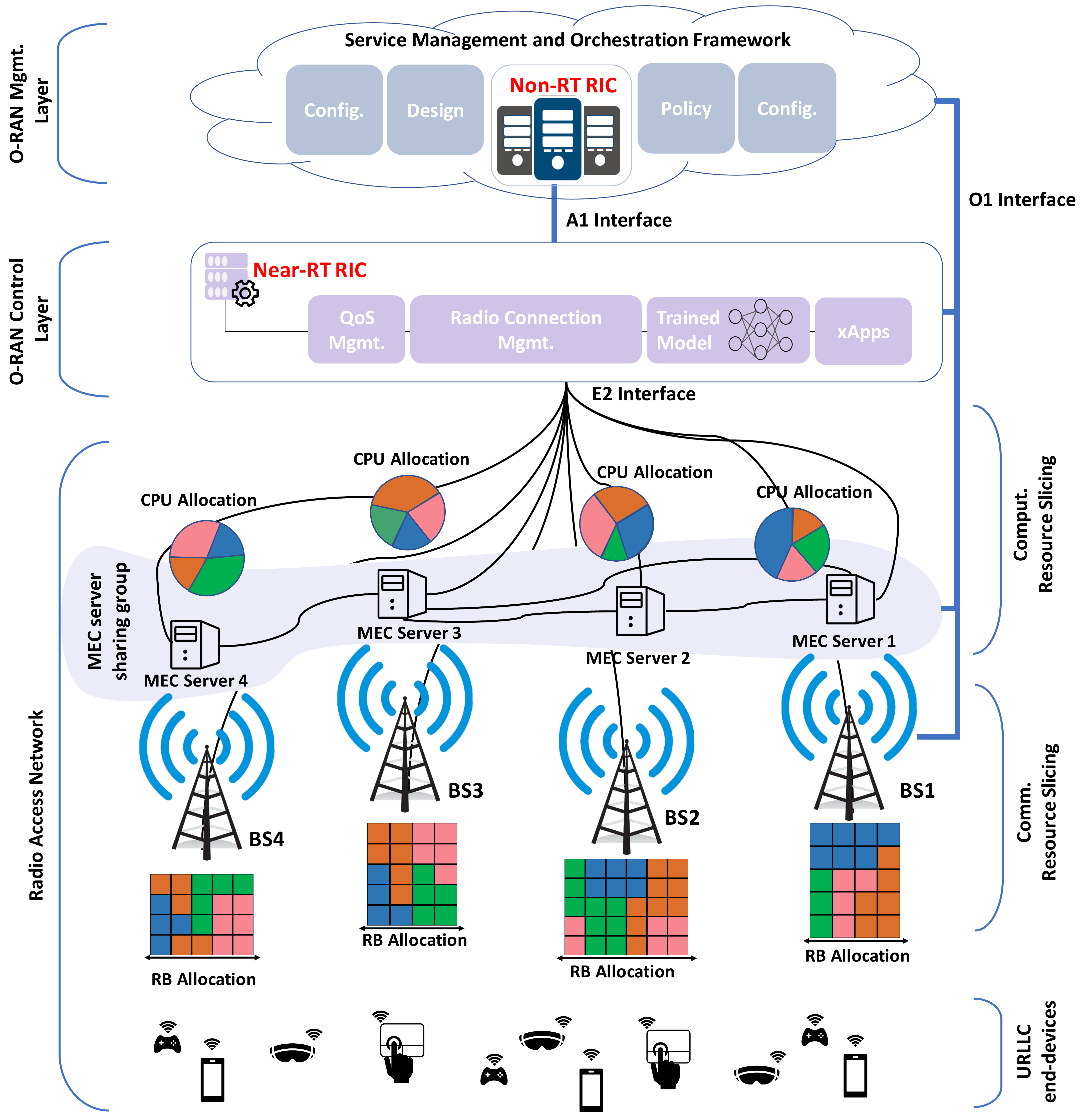}
	\caption{ Reference network RAN slicing model.}
	\label{fig:reference_network}
\end{figure}

\textbf{Network Model:}
We consider an O-RAN-based cellular network architecture, as depicted in Figure~\ref{fig:reference_network}, composed of 
\begin{enumerate*}[(i)]
    \item non-RT RIC that is directly connected to near-RT RIC through A1 interface, MEC servers, and gNodeBs through O1 interface to enables non-real-time control 
    of RAN elements and resources, 
    \item near-RT RIC that performs near-real-time control 
    of O-RAN elements and resources over the E2 interface,
    \item a set of MEC servers, controller by the near-RT RIC, 
    that form a MEC server sharing group, 
    \item a set of gNodeBs (gNBs) that provide communication resources to URLLC end-devices in their coverage area, and
    \item URLLC end-devices that offload their computing tasks under URLLC constraints, \ie strict latency, to the MEC servers.
\end{enumerate*}

Each gNB is attached to one MEC server to provide computation resources to URLLC end-devices. A gNB sharing group consists of a group of gNBs with highly overlapped in their communication coverage areas.
The communication and computation resources, considered in this work, are the resource block (RB) of gNB and the CPU core of the MEC server, respectively. RB is the smallest unit of radio resources that can be allocated to an end-device. 
A CPU core is defined as the computation capability in terms of CPU cycles per second.
We also consider the orthogonal frequency division multi-access offloading scenario, where the radio resources of a gNB are divided into multiple RBs. Hence, we avoid intra-cell interference where a specific RB is exclusively assigned to only one end-device.

\smallskip
\textbf{Assumptions:}
In our model, we consider the following assumptions:
\begin{enumerate*}[(i)]
    \item since radio resources are limited, it is challenging to provide enough orthogonal radio resources (in a multi-cell scenario). Thus, some gNBs can share the same radio resources, which may cause interference between cells. To counterbalance radio resources sharing and inter-cell interference reducing, the same set of radio resources can be assigned to multiple gNBs as long as the distance between them is sufficient to reduce inter-cell interference; 
    \item since a gNB sharing group is an area with strong overlaps between gNBs, the orthogonal resources are assigned to gNBs within one sharing group, which means the unavailability of interference within gNB sharing group; 
    \item each gNB covers a set of end-devices that are uniformly distributed in the gNB's coverage area; 
    \item each end-device is associated with only one gNB; and 
    \item each MEC server is equipped with multiple CPU cores to provide parallel computing.
\end{enumerate*}
The near-RT RIC performs the slicing operation of communication and computation RAN resources in two levels: 
\begin{enumerate*}[(a)]
    \item \textit{communication slicing level}, 
    and 
    \item \textit{computation slicing level} 
\end{enumerate*}

\smallskip
\textbf{Communication Slicing Level:}
Each gNB assigns a number of RBs to its associated end-devices. 
The RBs allocated to each end-device should ensure a low communication delay in offloading the task from end-device to the associated gNB through wireless transmission. A task's communication delay depends on its size and the total achievable data rate over the allocated RBs. 
Each end-device can be considered as an M/M/1 queuing system if:
\begin{enumerate*}[(i)]
    \item the arrival process of each end-device's tasks follows a Poisson distribution, and 
    \item the inter-arrival times of the tasks are independent and follow an exponential distribution.
\end{enumerate*}
Therefore, the delay experienced by a given task, in an offloading operation, can be calculated by applying Little's law.
%
We choose the M/M/1 queuing assumption since it is widely used to characterize wireless communication systems, especially in RAN slicing approaches. 
In addition, under different queuing assumptions, the mathematical analysis may be different and more complicated. Thus, for sake of simplicity, in this work we only assume the M/M/1 case.

\smallskip
\textbf{Computation Slicing Level:}
The computation resource slicing consists of allocating the required CPU cycles to successfully execute the offloaded tasks and meet the required QoS requirements. In fact, for each arrival task in each MEC server, the near-RT RIC needs to decide:
\begin{enumerate*}[(i)]
    \item where the task should be executed, and 
    \item how many computation resources should be allocated to this task.
\end{enumerate*}
For a given task, the near-RT RIC checks the available computation resources of the associated MEC servers, based on which it decides whether the task could be executed locally by its associated MEC server or forwarded to another MEC server in the same sharing group.
Then, the near-RT RIC allocates the required CPU cycles to execute this task. Indeed, checking the availability of computational resources in each MEC server should be performed by the near-RT RIC since it enables online information collection to optimize the control functions designed by the non-RT RIC. As a result, the near-RT RIC can take accurate computation slicing decisions based on the optimized control functions received from the non-RT RIC.

The computation delay can be defined as the ratio of the number of CPU cycles required to accomplish this task to the CPU cycles allocated by the near-RT RIC. When a task is forwarded to a different MEC server, the round-trip communication delay is added to the computation delay.


\subsection{Deep Reinforcement Learning based RAN Resource Slicing}
\begin{figure}[!t]
	\centering
	\includegraphics[width=1\linewidth]{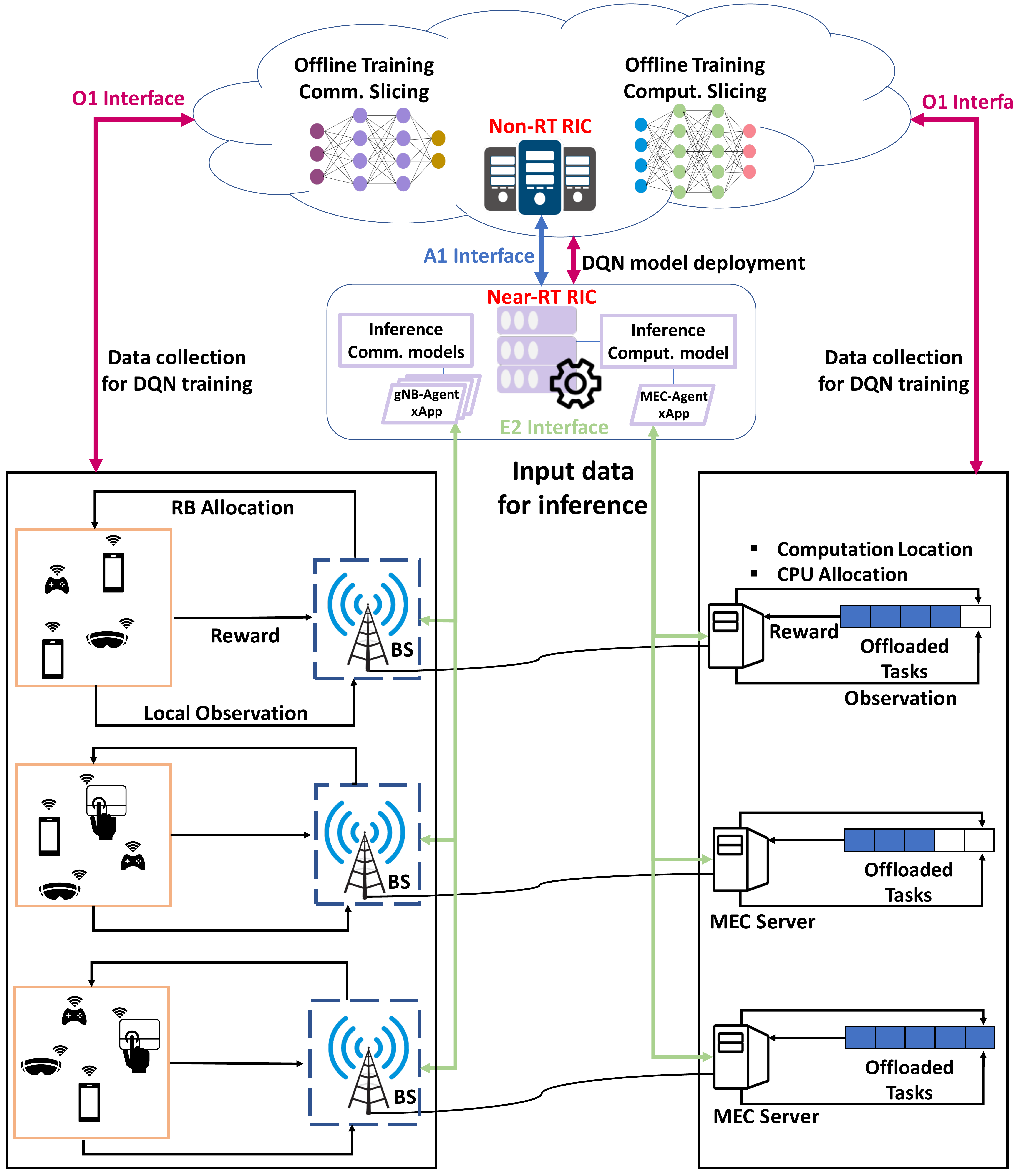}
	\caption{Deep Reinforcement Learning based RAN Resource Slicing.}
	\label{fig:drl_slicing}
\end{figure}

In an O-RAN architecture, the near-RT RIC is responsible for making resource allocation decisions. The efficiency of these decisions impacts the performance of the overall system. In particular, each gNB communicates the state of its environment, through the E2 interface, with the near-RT RIC that allocates the required RBs to the end-devices associated with this gNB. Similarly, the near-RT RIC collects information about the computation resource status of the MEC servers and allocates the CPU cycles, needed to execute the offloaded tasks in the appropriate MEC servers. 
In order to meet the requirements of URLLC services, communication and computation resource slicing need to be solved efficiently, especially in large-scale networks where the number of end-devices is huge. To overcome this challenge, DRL can be applied since it can efficiently deal with the curse of dimensionality problem. Figure~\ref{fig:drl_slicing} illustrates the overall working principle of the proposed DRL-based RAN resource slicing in an O-RAN architecture.

In this work, we opt for deep Q-learning (DQL) \cite{mnih2015human} to solve the RAN resource allocation problem in both communication and computation slicing.
DQL uses a deep neural network (DNN) as a Q-function approximator. This extension of the Q-learning algorithm is known as the deep Q-network (DQN) algorithm. Indeed, for a given input state, DQN generates a Q-value of all possible actions. 
The agent's experience is defined by the tuple (current state, action, reward, next state). Instead of immediately training the DNN by feeding it with successive experience tuples, the tuples are stored in a replay buffer according to the time sequence. During the DQN training process, the stored experiences are randomly sampled to train the DNN. The experience replay memory strategy allows efficient use of previous experiences in the DNN training process since it breaks the correlations in the observation sequences.

To further stabilize the approximation of the Q-value function, we employ the double DQN (DDQN) algorithm for both communication and computation slicing.
DDQN mitigates the overestimation problem that occurs in DQN algorithms since it applies a maximization operation on both the selection and evaluation actions. Specifically, DDQN uses two neural networks: main Q-network to select action, and target Q-network to calculate the estimated Q-value of each selected action. The main Q-network is trained by minimizing the loss function. The latter calculates the mean square error between the current Q-values of actions selected by the main Q-network and their estimated Q-values calculated by the target Q-network.
%

In the O-RAN architecture, Figure~\ref{fig:drl_slicing}, we consider that each gNB is controlled by a DRL agent, called gNB-agent, which performs the communication resource slicing between the associated end-devices of this gNB. For the computation slicing level, we consider that the MEC server sharing group is controlled by a DRL agent, called MEC-agent, which allocates the CPU cycles required to successfully execute the offloaded tasks. Each DRL agent runs in a xApp on the near-RT RIC and manages its resources through the E2 interface.

Before describing the proposed DRL-based approach, we first model each resource slicing problem  as a Markov decision process (MDP). 

\smallskip
\textbf{MDP-based Communication Resource Slicing:}
Each gNB-agent observes its environment and allocates RBs to its associated end-devices. For each gNB, the communication resource slicing is modeled as a single-agent MDP. 

\begin{itemize}
    \item \textit{The State Space:} 
    The state space of a gNB-agent includes the: (i) set of associated end-devices, (ii) 
    available RBs, 
    (iii) channel gain between the gNB and its associated end-devices 
    and (iv) maximum delay threshold required by the URLLC service.

    \item \textit{The Action Space:} 
    A gNB-agent has to decide which RBs should be allocated to each of the associated end-devices. Since an end-device can have more than one RB to meet the desired QoS, 
    an action is defined by a row vector where each element represents the RB - end-device assignment.

    \item \textit{The Reward Function:} 
    The reward received by the gNB-agent depends on whether it successfully allocated the required RBs or not. An action is considered to be successful if it meets the constraints of the RB allocation model. Since our objective is to minimize the communication delay, the received reward is the inverse of the sum of all communication delays of all tasks offloaded by the associated end-devices. Otherwise, the gNB-agent is penalized with a negative reward.
\end{itemize}

\smallskip
\textbf{MDP-based Computation Resource Slicing:}
We model the computation resource slicing as a single-agent MDP.

\begin{itemize}
    \item \textit{The State Space:}
    The state space of the MEC-agent is given by information about each MEC server including the offloaded tasks and the available computation resources. Since the observed state is unknown directly to the MEC-agent, each MEC server regularly updates the MEC-agent about its local state. An update can include task-related information such as the number of tasks currently in its buffer, the size of each task, the number of CPU cycles needed, and a maximum delay threshold required by the URLLC service.

    \item \textit{The Action Space:}
    The MEC-agent 
    decides the computation resource allocation for each offloaded task. A decision includes: (i) in which MEC server a task should be executed 
    , and (ii) CPU cycles allocation that consists in determining the number of CPU cores to be assigned for computing a received task.

    \item \textit{The Reward Function:}
    The reward obtained by the MEC-agent 
    depends on whether the chosen action is feasible or not and at what level the computation delay was minimized. An action is considered feasible if it meets the computation resource allocation constraints 
    The received reward is the inverse of the sum of computation delays of all tasks offloaded by the end-devices. Otherwise, the received reward is set to a negative value to prevent the MEC-agent from choosing non-feasible actions in the future.
\end{itemize}

\subsection{Deep Q-learning Slicing Algorithm}
A DDQL-based approach consists of two main phases: the training phase and the implementation phase, \ie inference. In the training phase, a DDQN is trained in an offline manner. In the implementation phase, the agent takes actions in an online manner based on its trained DDQN. In the O-RAN architecture, the DDQN model is trained offline in the non-RT RIC, while the model inference is deployed in the near-RT RIC. The non-RT RIC uses the O1 interface to collect data for offline model training. Note that the trained model can undergo an evaluation step validating that it is reliable for deployment in the near-RT RIC. The model inference is executed and fed with online data, through the E2 interface, to produce the slicing actions that will be used in the resource allocation operation.
The training and implementation phases of both slicing levels are conducted in the same way\footnote{The term agent is used to refer to the gNB-agent or MEC-agent, based on the slicing level.}, which can be summarized as follows.

\begin{figure*}[t]
    \centering
        \subfloat[Training reward.]{
            \includegraphics[width=0.9\columnwidth]{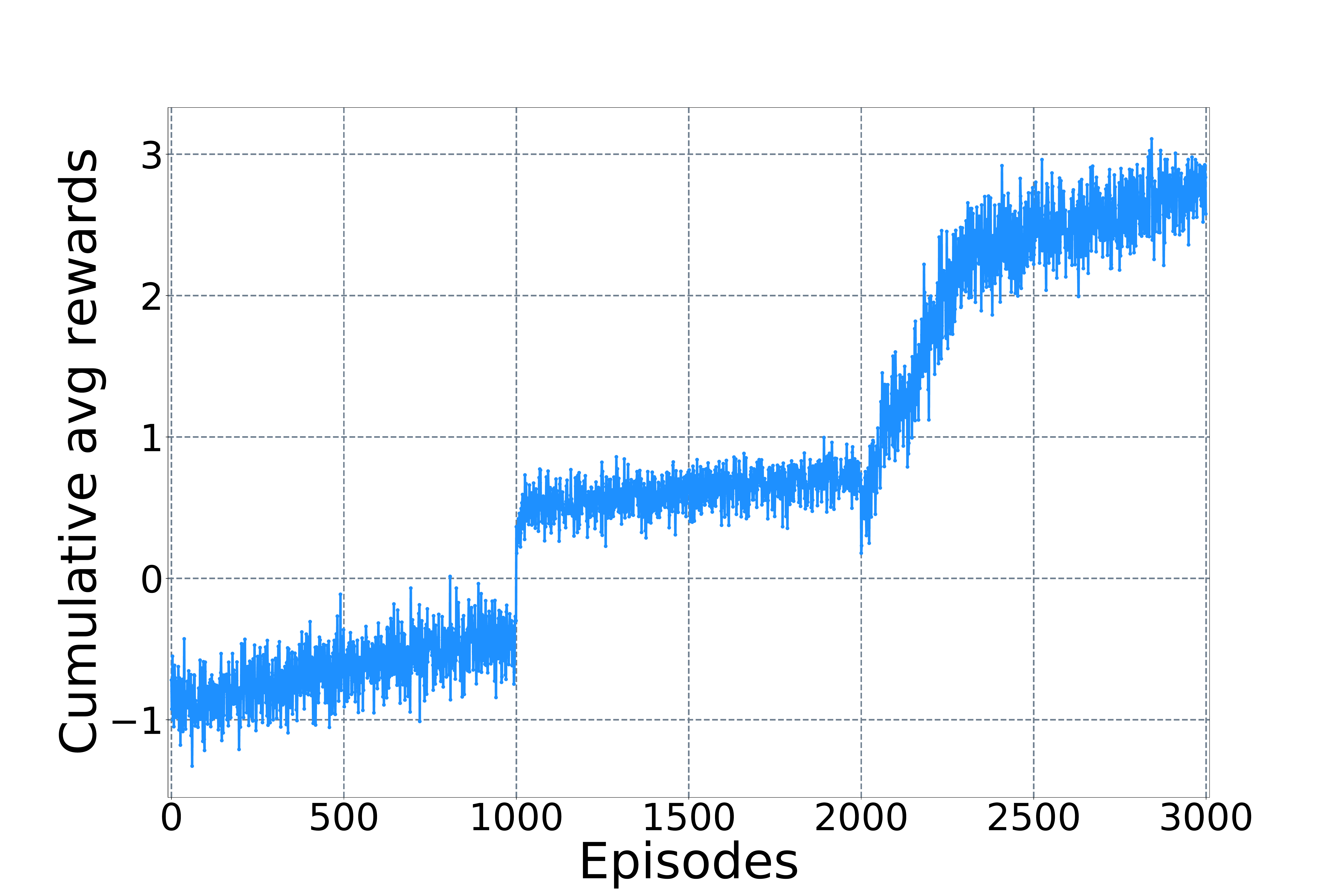}
            \label{fig:TR_Comm}
        }
        \quad
        \subfloat[Training loss.]{
            \includegraphics[width=0.9\columnwidth]{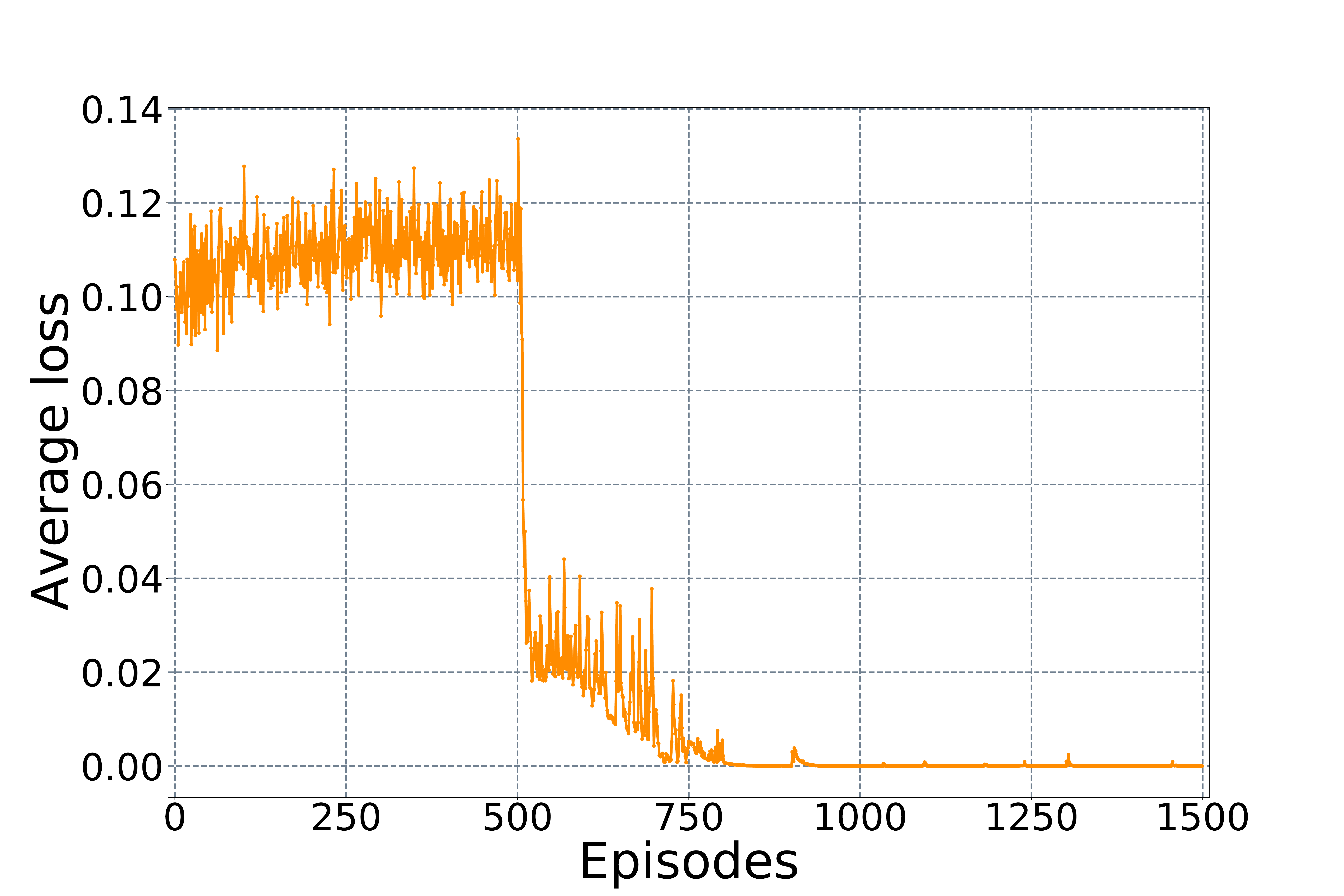}
            \label{fig:TL_Comm}
        }
    \caption{Training performance of the communication model.}
    \label{fig:T_Comm}
\end{figure*}

\begin{figure*}[t]
    \centering
        \subfloat[Training reward of the computation model.]{
            \includegraphics[width=0.9\columnwidth]{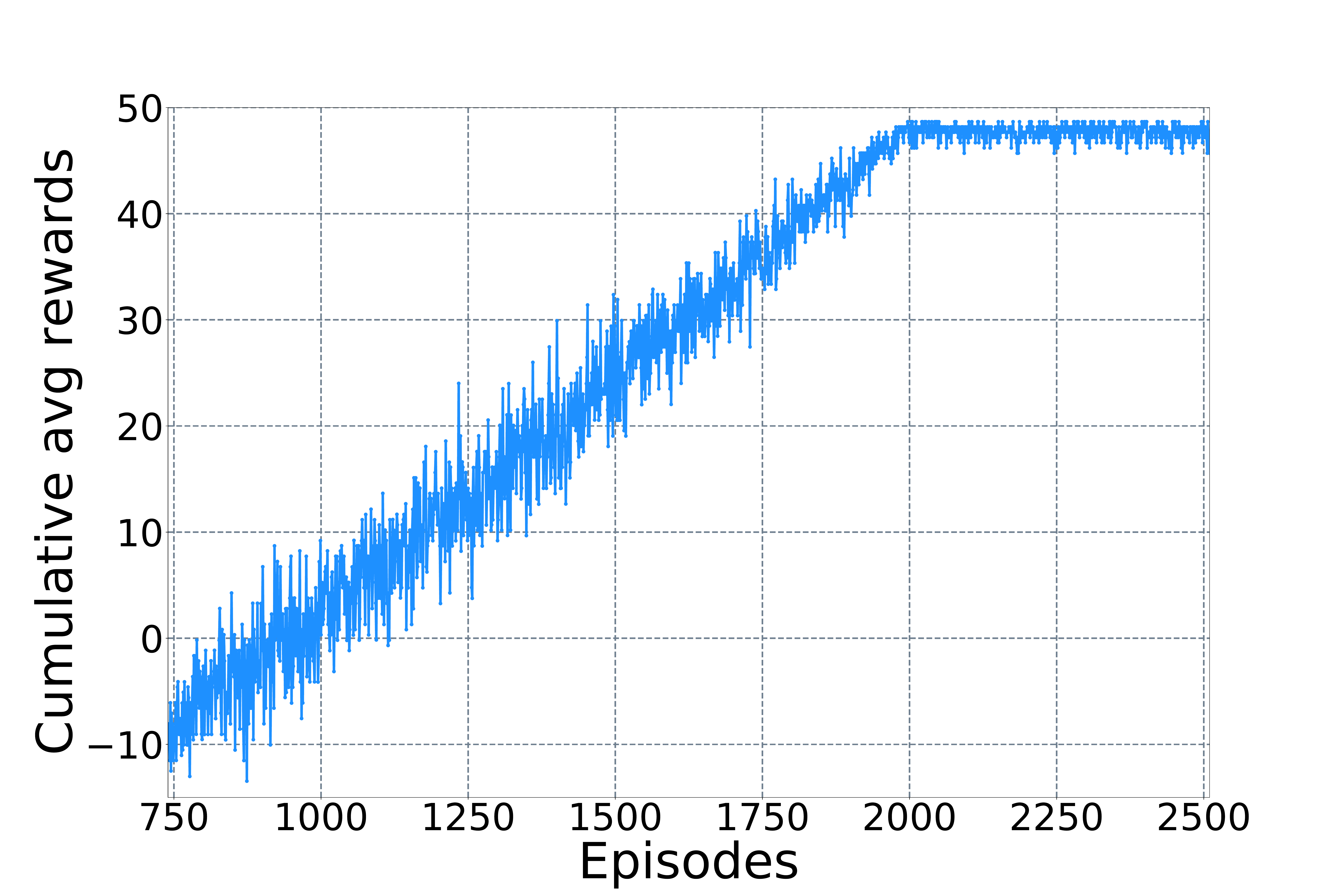}
            \label{fig:TR_Comp}
        }
        \quad
        \subfloat[Training loss of the computation model.]{
            \includegraphics[width=0.9\columnwidth]{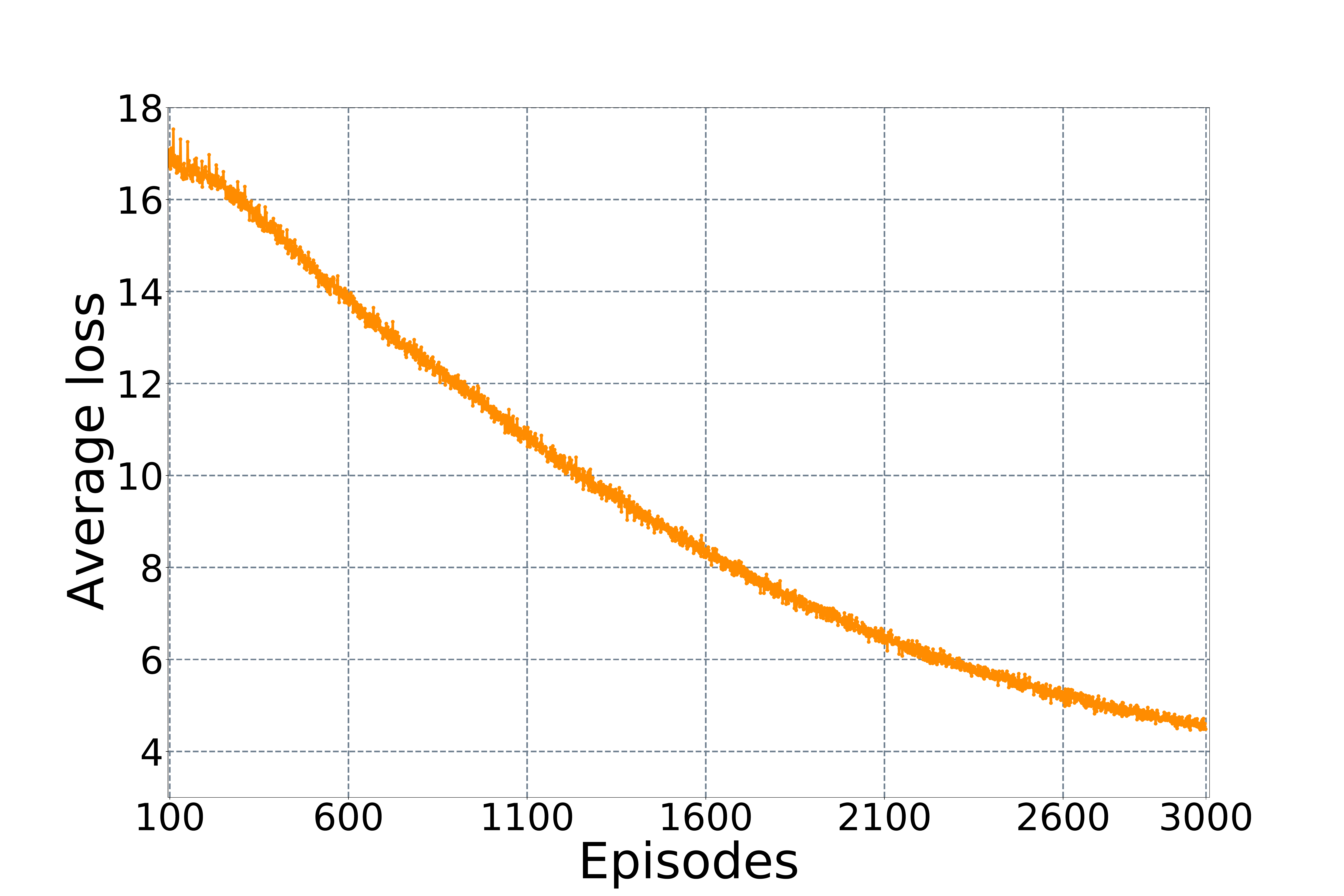}
            \label{fig:TL_Comp}
        }
    \caption{Training performance of the computation model.}
    \label{fig:T_Comp}
\end{figure*}

\smallskip
\textbf{The Training Phase:}
It consists of several episodes and requires, in each episode, the state of the environment as input. As output, a trained DDQN is produced. 

The learning process begins by iterating the episodes. At the beginning of each step for each episode, the agent observes the state of its environment and chooses an action according to an $\epsilon$-greedy policy. With the help of $\epsilon$-greedy policy, the training process is balanced between exploitation and exploration. At each step, the agent takes a random action with a probability of $\epsilon$ 
and follows its current policy by choosing the action with the highest Q-value in the remaining time. 
As the training process proceeds, the $\epsilon$ value gradually decreases, indicating that the agent becomes more confident to optimally interact with the environment and choosing optimal actions. 
The obtained experience tuple is stored in a replay buffer. When the buffer contains enough experiences, the agent picks a random sample to create training data. Then, it performs the gradient descent algorithm to minimize the loss function and update the parameters of the main Q-network. On the other hand, the target Q-network parameters do not need to be updated at each training step but replaced by the main Q-network parameters with a certain frequency.

\smallskip
\textbf{The Implementation Phase:}
Once the offline training phase is complete, the agents can use their trained DDQNs to efficiently allocate RBs and CPU cycles. During the implementation phase, when a new state of the environment is observed, the agent selects 
the action with the highest Q-value. Afterward, end-devices can offload their tasks to the associated gNB using the optimal RBs. Then, tasks will be executed by the MEC servers using an optimal CPU cycle allocation.

\section{Performance Evaluation}
\label{sec:evaluation}

    \begin{figure*}[t]
    \centering
        \subfloat[Communication model.]{\includegraphics[width=0.9\columnwidth]{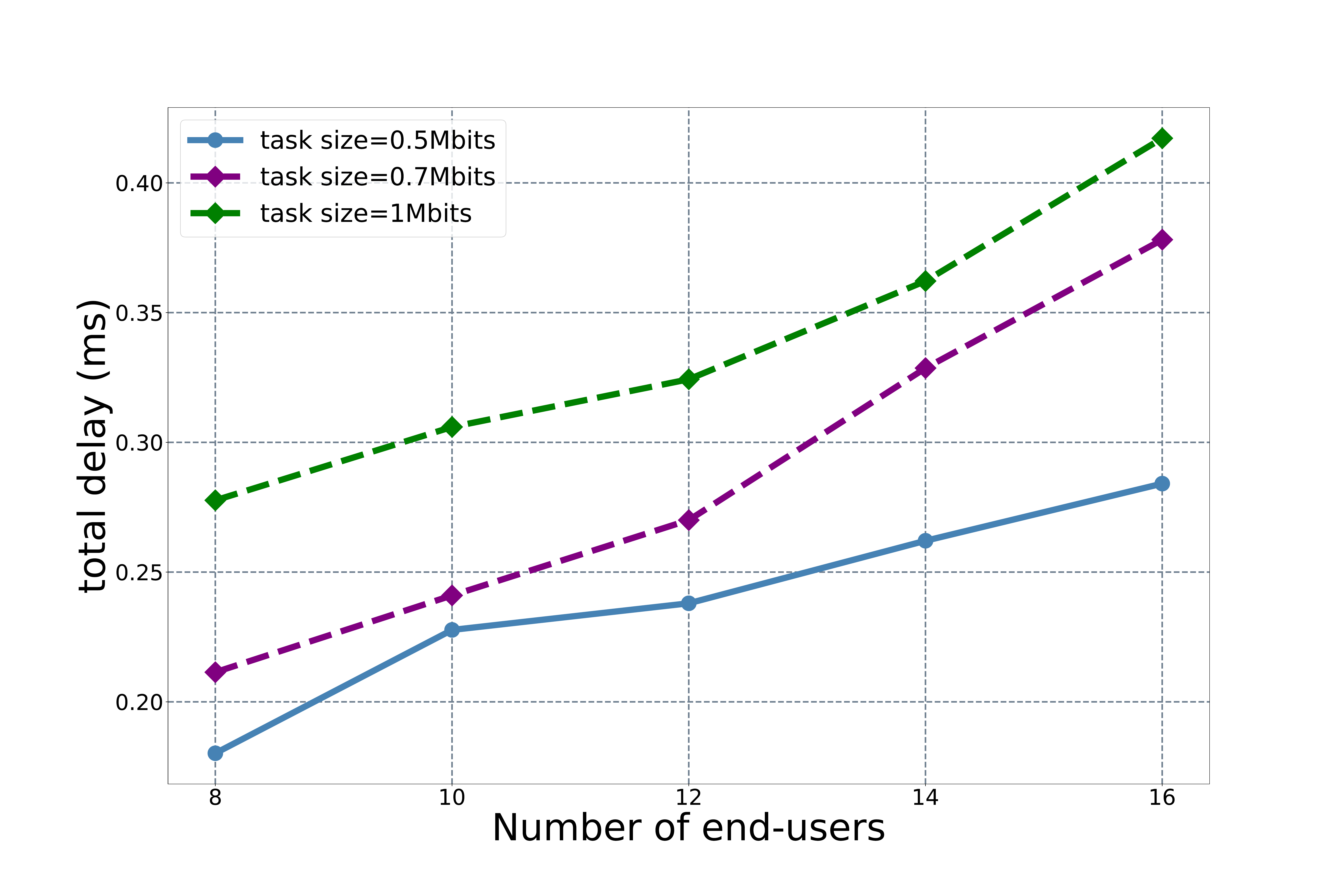}\label{fig:D_Comm}}
        \quad
        \subfloat[Computation model.]{\includegraphics[width=0.9\columnwidth]{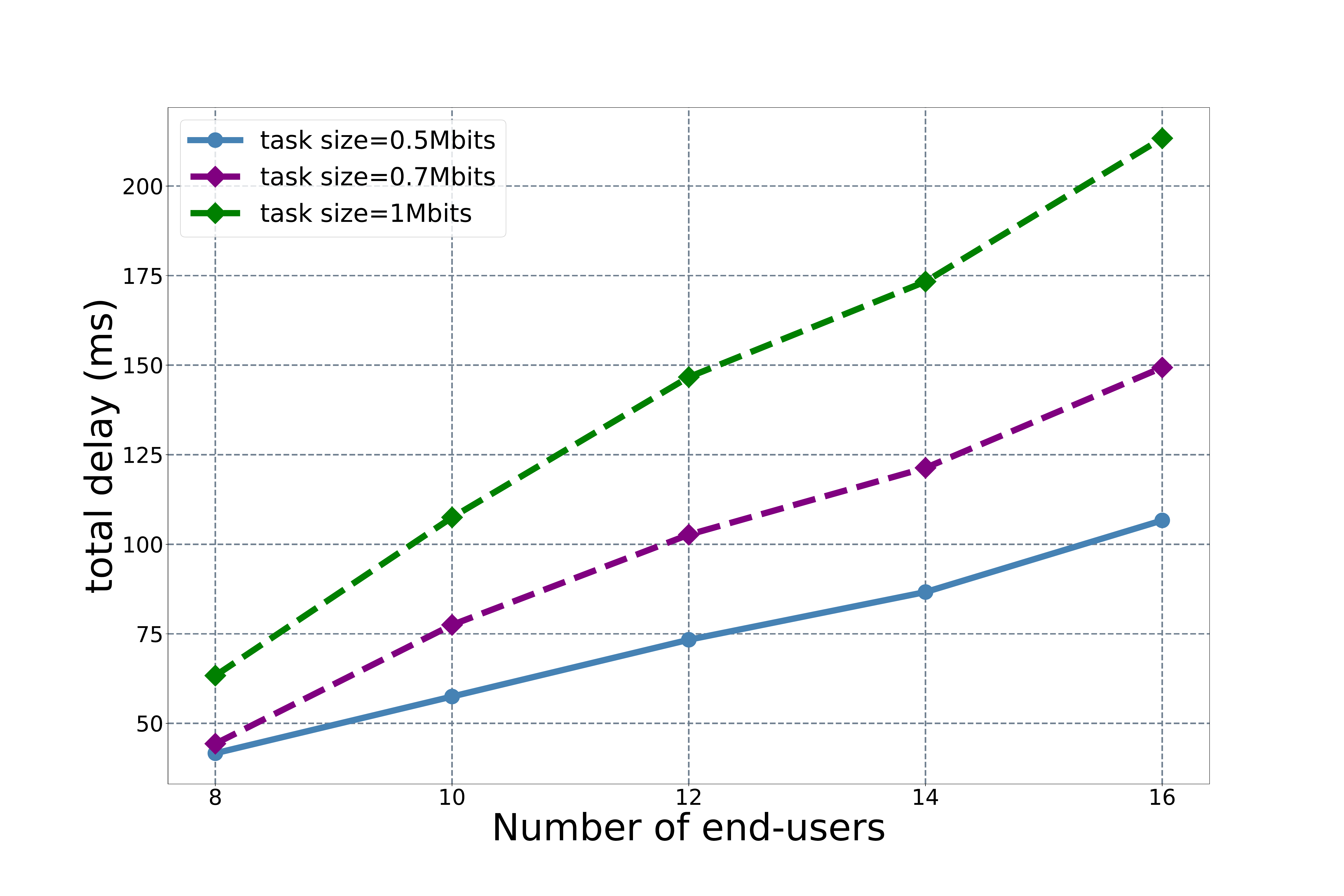}\label{fig:D_Comp}}
    \caption{Delay performance.}
    \label{fig:Delay}
\end{figure*}

\smallskip
\textbf{Simulation Setup and Scenario:}
Following the reference network model shown in Figure~\ref{fig:reference_network}, we implemented an O-RAN-based cellular network architecture with four gNBs using Python programming language.
The gNBs are deployed in a geographical zone modeled by a square of a side of 2000 m. Each gNB covers a circular area with a radius of 500 m and is accompanied by one MEC server. Each MEC server is equipped with four CPU cores with a computation capability equals to 3 gigacycles each. End-devices, with URLLC services, are uniformly distributed within the coverage area. Each end-device is associated with only one gNB and can offload only one task at a time.
The data size of each task is uniformly distributed from 0.5 MB to 2 MB and the required CPU cycles to compute one bit is 400. The transmission power of end-devices is 23 dBm, while the bandwidth of an RB is 180 kHz and the noise power is -114 dBm.
The DDQNs were implemented and trained using the PyTorch framework, an open source machine learning library written in Python. For the training, we used two fully connected hidden layers composed of 256 neurons each, ReLu as the activation function, Adam as the optimizer, and the mean square error as the loss function. 
The learning rate of the communication model and the computation model is 0.01 and 0.001, respectively, while the mini-batch size is 64 and 256, respectively.
The simulations were conducted on a laptop with a 2.2 GHz Intel i7 Processor, 16 GB of RAM, and NVIDIA GeForce GTX 1070 graphic card.
%

\smallskip
\textbf{DDQN Training Performance:}
Figures~\ref{fig:T_Comm}~and~\ref{fig:T_Comp} show the training performance of the communication and the computation models, respectively. Indeed, Figures~\ref{fig:TR_Comm}~and~\ref{fig:TR_Comp} illustrate the convergence of the communication and computation DQL algorithms, respectively, versus training episodes. 
It can be seen from both figures that when the number of training episodes increases, the cumulative average reward grows. We also notice that the convergence of the computation DQL algorithm is faster (converge after 2000 episodes) than that of the communication DQL algorithm (from episode number 2500). The convergence of the communication DQL algorithm is relatively slow due to the mobility of end-devices, so the channel gains between the gNBs and the end-devices change frequently. These convergence results demonstrate the effectiveness of the proposed algorithms. 
 Indeed, when end-device mobility is high, the wireless channel changes rapidly, which impacts the accuracy of the channel state information (CSI) at the gNB. A gNBs estimates the gain between itself and its associated end-devices using the CSI. In the communication slicing level, the estimated gain is used as input for the DDQN algorithm during its training phase. Therefore, an imperfect CSI, caused by the high mobility of end-devices, can reduce the learning performance of the proposed DDQN algorithm. As a result, under the imperfect CSI constraint, the DDQN algorithm takes more time to accurately learn the appropriate policies. However, once the offline training of the DDQN algorithm is performed, the learned policy can be applied rapidly to obtain the resource allocation solution.

Figures~\ref{fig:TL_Comm}~and~\ref{fig:TL_Comp} show how the behavior of the loss function for both communication and the computation DQL algorithms, respectively, evolves as training proceeds. In the early stages of the training process, the performance of both algorithms is weak due to exploration phenomena, \ie the gNB-agents and the MEC-agent take random actions more than exploiting what they have learned. The loss value decreases to reach a minimum value at the end of the training process, which indicates that the Q-value approximation has become accurate.

\smallskip
\textbf{Delay Performance:}
In this experiment, we evaluated the performance of the proposed RAN slicing approach in terms of the delay experienced by tasks. For each resource slicing model, 
we varied the number of end-device and calculated the delay experienced by tasks for different task sizes, \eg 0.5 MB, 0.7 MB, and 1 MB. Based on the observed results in Figures~\ref{fig:D_Comm}~and~\ref{fig:D_Comp}, we make the following observations: (1) it is clear that as the number of end-device increases, the delay experienced by the tasks increases, and (2) the performance gap between the three task sizes remains relatively constant for a different number of end-devices. For (1) when the number of the end-devices becomes higher, the competitiveness among end-devices increases to obtain sufficient RBs and CPU cycles. 
In fact, when the number of the end-devices is low, the gNB-agents and the MEC-agent can assign several RBs and CPU cycles, respectively, to only one end-device. In contrast, when the number of end-devices is high, the gNB-agents and the MEC-agent, respectively, assign a minimum of RBs and CPU cycles to satisfy all end-devices. Observation (2) demonstrates the scalability of the proposed RAN slicing approach under a dense network topology.

\section{Conclusion and Future Work}
\label{sec:conclusion}
In this article, we designed a two-level RAN slicing approach to allocate communication and computation resources to URLLC end-devices. The approach is integrated in the O-RAN architecture with MEC technology. We modeled each RAN resource slicing problem as a single-agent MDP. Then, we developed a DQL algorithm to solve each resource slicing problem and described the role of non-RT and near-RT RICs in performing slicing operations. The proposed DQL-based solution shows robust and efficient performance in meeting the requirements of URLLC services.
The results of this study show that a deep reinforcement learning based RAN resource slicing architecture such as the one presented is promising and deserves further investigation.
 However, it is also clear that this flexibility may come with the price of possible slice specification violations. One direction that would be worth investigating is the relationship between reconfiguration interval duration, the probability of slice specification violations, DLR training accuracy, and retraining frequency. Another interesting issue to study is the trade-off between the penalties of overbooking strategies and slice requirement violations in a DRL-based resource partitioning architecture.

\section*{Acknowledgments}
The authors would like to thank the Natural Sciences and Engineering Research Council of Canada, for the financial support of this research.

\bibliographystyle{IEEEtran}
\bibliography{Ref}

\end{document}